\documentclass[epj,draft]{svjour}
\input{epsf}
\begin{document}
\title{Two interacting particles in a disordered chain III:\\
Dynamical aspects of the interplay Disorder - Interaction}
\titlerunning{TIP III: Dynamical aspects of the interplay Disorder - Interaction}
 
\author{Samuel De Toro Arias\inst{1,2}\and Xavier Waintal\inst{1}
\and Jean-Louis  Pichard\inst{1,}\thanks{\email{pichard@spec.saclay.cea.fr}}}
\institute{CEA, Service de Physique de l'Etat Condens\'e, 
           Centre d'Etudes de Saclay, F-91191 Gif-sur-Yvette, France \and 
	   Laboratoire de Physique de la Mati\`{e}re Condens\'{e}e,
           CNRS UMR $6622$, Universit\'{e} de Nice-Sophia Antipolis,
           Parc Valrose,\\ B.P. $71$, $06108$ Nice Cedex $2$, France.}


%
\abstract{ 
The interplay between the quantum interferences responsible for one 
particle localization over a length $L_1$, and the partial dephasing 
induced by a local interaction of strength  $U$ with another particle 
leading to partial delocalization over a length $L_2 > L_1$, is 
illustrated by a study of the motion of two particles put close to each 
other at the initial time. Localization is reached in two steps. First, before 
the time $t_1$ necessary to propagate over $L_1$, the interaction slows 
down the ballistic motion. On the contrary, after $t_1$ the interaction 
favors a very slow delocalization, characterized by a $\log(t)$ spreading 
of the center of mass, until $L_2$ is reached. This slow motion is related 
to the absence of quantum chaos in this one dimensional model, the 
interaction being only able to induce weaker chaos with critical spectral 
statistics. Under appropriate initial conditions, the motion remains 
invariant under the duality transformation mapping the behavior at small 
$U$ onto the behavior at large $U$.
}
\maketitle

\section{Introduction}
 
 In this third work of a series~\cite{wp,wwp,ww} concerning two 
interacting particles (TIP) in a one dimensional random lattice, 
we use the time dependent Schr\"odinger equation 
to describe the competition between the one particle 
quantum interferences induced by the random potential 
leading to localization, and the mutual dephasing induced by a 
local interaction and leading to partial delocalization. 
At the initial time, a wave packet representing two particles in 
two neighboring sites is constructed in the middle of a disordered 
chain  of size $L$. A repulsive on site interaction of strength $U$ 
is considered. The particles are assumed to be two electrons with 
opposite spins, the initial wave function is symmetric and remains 
symmetric during the quantum motion. We study the short times where 
the particles visit scales small compared to the one particle 
localization length $L_1$ till the long times where the spreading 
of the center of mass saturates at the TIP localization length 
$L_2 \gg L_1$. For this, we use an efficient automaton-like 
algorithm adapted to discrete scalar wave propagation in a system of 
finite size $L$.  We restrict the study to strong localization, such 
that the quantum motion is the same when the size increases from  
$L= 512$ to $L=1024$ and $t \rightarrow \infty$. This guarantees that 
our conclusions are not biased by finite size effects coming from 
successive boundary reflexions. The price to pay for this is to 
consider relatively small values for $L_1$ and $L_2$. 

 The TIP-dynamics is characterized by two times $t_1$ and $t_2$,  
where a scale of order of $L_1$ and $L_2$ is respectively explored. 
Between $t_1$ and $t_2$, we find that the spreading of the center 
of mass is extremely slow. To give an order of magnitude, $L_1 
\approx 16 $ can be quickly reached after $t_1 \approx 100$ (in 
units of time)  while a time $t_2 \approx 10^4$ is necessary to 
reach only $L_2 \approx 2 L_1$. In this regime of interaction 
assisted propagation, we find that the center of 
mass spreads with a $\log(t)$-law, quite different from a 
previously assumed diffusion law. This is consistent with the 
observation~\cite{wwp} that the interaction can never drive the 
TIP system in one dimension to full quantum chaos with Wigner-Dyson 
spectral statistics. Only a weak critical chaos can be established, 
where the spectral fluctuations are statistically similar to those 
characterizing~\cite{bogomolny} different critical one particle 
spectra ($3d$ Anderson model at the mobility edge~\cite{shapiro}  
or certain pseudo-integrable billiards).  Another important phenomenon 
illustrated by this study is the inversion of the effect 
of the interaction when the ballistic motion ($t<t_1$) becomes 
sub-diffusive ($t_1<t<t_2$): first $U$ defavors the ballistic 
propagation before having an opposite delocalizing effect.

The paper is organized as follows. The model and the used algorithm are 
introduced (Section \ref{model}).  Then a series of useful results for 
understanding the complex features of TIP dynamics (Section \ref{Rap}) 
are shortly summarized. Illustrations of the TIP delocalization phenomenon 
are given when $t \rightarrow \infty$. The role of the finite size 
effects are studied for $L=512$, and the values of $L_1$ and $L_2$ where they 
can be neglected are estimated (Section \ref{L2}). In the remaining part, 
we study how is reached this long time limit, after two successive regimes 
of the quantum motion. First, a ballistic regime ($t<t_1$) is 
studied where $U$ defavors propagation (Section \ref{ballistic}). 
The dependence of the TIP dynamics on the chosen initial wave packet is 
illustrated. We consider both two particles put at $t=0$ on the same site 
(energy $\approx U$) and on two neighboring sites (energy $\approx 0$). 
Following the initial condition, the dynamics probes two different sets 
of states in the large $U$-limit: molecular states of energy $\approx U$ 
and hard core boson states of energy $\approx0$, as discussed in 
Ref.~\cite{wwp}. The duality transformation mapping 
the behavior at small $U$ onto the behavior at large $U$ is illustrated 
for appropriate initial conditions. A study of $t_1$ follows 
(Section \ref{t_1}) before describing the sub-diffusive regime 
where interaction favors a slow TIP propagation ($t_1 < t < t_2$: 
Section \ref{sousdif}). The $\log(t)$ spreading of the center of mass 
is related to the long life-time observed in Ref.~\cite{wp} for the 
free boson states (TIP states for $U=0$). The relation between the 
observed very slow delocalization, the multifractal measure~\cite{wp} 
characterizing the interaction induced hopping terms coupling the free 
boson states, and the critical weak chaos observed in Ref.~\cite{wwp} are 
discussed. 

\section{Model and numerical algorithm} \label{model}

 To study the motion of two electrons with opposite spins in a one 
dimensional Anderson tight binding model with on site interaction, 
we have to numerically solve the discretized TIP Schr\"odinger equation: 
\begin{equation}
 \frac{i}{2 \epsilon} \left( |\psi(t+\epsilon) \rangle - |\psi(t-\epsilon) 
\rangle \right) 
= {\cal H} |\psi(t) \rangle
\label{anderson},
\end{equation}
where $ {\cal H}= H_0 \otimes 1 + 1 \otimes H_0 + H_{\rm int} $. 
$H_0$ is the one particle tight binding Anderson Hamiltonian:
$$ 
H_0 = \sum_{n=1}^{L-1} \left( |n\rangle \langle n+1|+ 
|n+1\rangle \langle n| \right) + \sum_{n=1}^L V_n |n\rangle 
\langle n|.
$$
$L$ is the system size and $V_n$ are independent random variables, 
uniformly distributed inside the interval $[-W,W]$. The ket 
$|n\rangle$ stands for the electronic orbital located at  
the site $n$ of the one dimensional lattice. The eigenstates 
$|\alpha\rangle$ of $H_0$ are localized on a length $L_1$ (with 
$L_1 \approx 24/W^2$ at the band center). $H_{\rm int}$ is the on 
site interaction:
$$ H_{\rm int} = U \sum_{n=1}^L |n\rangle \otimes |n\rangle 
\langle n| \otimes \langle n|.
$$
In what follows, all the lengths are given in lattice spacing units and 
all the energies in units of $1/\epsilon$. The times are then expressed 
in the corresponding units ($\epsilon$). The sites are labelled from 
$-L/2$ to $+L/2$, such that the site $n=0$ is located in the middle 
of the chain. The initial condition corresponds to 
$$\psi_{n_{1},n_{2}}(0)=\psi_{n_{1},n_{2}}(\epsilon) =
\frac{1}{\sqrt{2}}(\delta_{n_1,0}\delta_{n_2,\rho_0} + \delta_{n_2,0}
\delta_{n_1,\rho_0}),$$
where $\langle n_1 n_2|\psi(t)\rangle=\psi_{n_{1},n_{2}}(t)$.

 When $\epsilon$ is small enough, the discrete time Eq.(\ref{anderson}) 
has the same physical content as its continuous version.

To solve Eq.(\ref{anderson}), we use an automaton-like algorithm~\cite{detoro}, 
which relies on a formulation of discrete scalar wave propagation 
in an arbitrary inhomogeneous medium by the use of elementary processes 
obeying a discrete Huygens' principle and satisfying fundamental symmetries,
as described in Ref.~\cite{detoro}. Our algorithm
avoids the direct discretisation procedure and incorporates
the symmetries underlying the Anderson model at the lowest
stage of the construction. As a consequence the algorithm preserves the 
unitarity of the dynamics, insuring the normalization of the wavefunction 
at all times, $\sum_{n_{1},n_{2}} |\psi_{n_{1},n_{2}}(t)|^2=1$, 
up to a small correction of order $\epsilon^2$. Besides, the construction 
is optimized for implementing the algorithm on massively parallel machines. 
The numerical simulations have been carried for a time step 
$\epsilon=0.05$. The simulations were performed on a $16$K processor 
Connexion Machine. Given a value of the disorder strength $W$ and a 
disorder configuration, the wavefunction has been calculated for chains 
of length as large as $L=1024$ and up to a maximum of $10^6$ units of
time.

\section{ Review of some useful results} \label{Rap}

   Since the two body quantum motion is quite complex, it is useful to 
have in mind a few previous results that we shortly summarize. 

\subsection{Non linear $\sigma$ model}\label{a} 

  The first analytical descriptions~\cite{s,fmgp} of the TIP system 
 are mainly based on simplified random matrix Hamiltonians with 
 independent Gaussian entries. The purpose was first 
 to explain the new phenomenon of pair propagation 
 at scales larger than $L_1$ and the pair localization at a scale 
 $L_2 \gg L_1$.  In Ref.~\cite{fmgp}, from such an effective random matrix 
 Hamiltonian, a supersymmetric nonlinear $\sigma$ model 
 was derived, closely related with the one found by Efetov 
 for non interacting electrons in disordered metals. The approach was 
 mainly developed for an arbitrary dimension $d$. We recall the 
 conclusions for a strictly one dimensional model ($d=1$).

  Let us denote by $|\alpha \rangle$  a one particle state located near 
 the site $n_{\alpha}$, $|\alpha\beta\rangle$ the symmetrized  product 
 states forming the eigenbasis of the TIP Hamiltonian when $U=0$. 
 Since they are the symmetric eigenstates without interaction, we call 
 them ``free boson states'' as in Ref.~\cite{wwp,ww}. We denote by
 $\Gamma_{\alpha\beta}$ their inverse lifetime when the interaction 
 is switched on. The TIP  supersymmetric $\sigma$ model gives 
 $\Gamma_{\alpha\beta}$ in agreement with the Fermi golden rule:
 \begin{equation} 
 \Gamma_{\alpha\beta}  \sim 2\pi 
 |\langle\alpha\beta|H_{\rm int}|\alpha\beta\rangle|^2 \nu_{\rm eff} 
 \approx \Gamma_1  f\left(\frac{ |n_{\alpha}-n_{\beta}|}{L_1}\right),
 \end{equation} 
 where 
 \begin{equation}
 \Gamma_1 \approx 2\pi \frac{U^2}{BL_1}.
 \end{equation}
 $B$ is the kinetic energy (band width), 
 $\langle\alpha\beta|H_{int}|\gamma\delta\rangle$ 
 the typical interaction matrix element between two free boson states 
 and $\nu_{\rm eff}$ the density of free boson states $|\gamma\delta\rangle$ 
 coupled to $|\alpha\beta\rangle$ by the interaction. 
 $ f(x) \approx 1 $ when $x <1$, i.e. when the typical distance 
 $|n_{\alpha}-n_{\beta}|$ between the localized states $|\alpha\rangle$ 
 and $|\beta\rangle$  is smaller than $L_1$ and $f(x) \approx \exp(-x)$
  when $| n_{\alpha}-n_{\beta}| > L_1$. The lifetime of 
 $|\alpha\beta\rangle$ depends on the spatial overlap between $|\alpha\rangle$ 
 and $|\beta\rangle$. Localized at a distance $|n_{\alpha}-n_{\beta}|$ large 
 compared to $L_1$, the two particles have an exponentially small probability 
 to be on the same site and to feel the on site interaction. The corresponding 
 lifetime is  exponentially large. For $|\alpha\rangle$ and $|\beta\rangle$ 
 localized inside  the same localization domain of size $L_1$, the lifetime 
 $\Gamma_1^{-1}$  defines an important characteristic 
 time. In Ref.~\cite{fmgp}, $\Gamma_1^{-1}$ and $L_1$  are respectively  the 
 smallest resolved time and length scales. The main assumption of the 
 approach, contained in the effective random  matrix Hamiltonian from which 
 the $\sigma$ model is derived, is that the one particle states are 
 essentially ergodic, chaotic and random inside their localization  domain. 
 This is a simplification of the one body dynamics inside $L_1$, which 
 amounts to assume for the single particle a ``zero-dimensional'' dynamics 
 inside $L_1$. For $L=L_1$, one gets 
 $\langle\alpha\beta|H_{int}|\gamma\delta\rangle \approx \pm U/L_1^{3/2}$ 
 and $\nu_{\rm eff} \approx L_1^2/B$ near the band center, and eventually  
 the above expression for $\Gamma_1$. As far as the dynamics are concerned, 
 the main conclusions of Ref.~\cite{fmgp} are as follows. For 
 $t > \Gamma_1^{-1}$, the time where the pair size $\rho (t)$ is roughly equal
 to $|n_{\alpha}-n_{\beta}|$ can be estimated from the lifetime of the 
 free boson state $|\alpha\beta\rangle$ with $|n_{\alpha}-n_{\beta}| 
 \approx \rho(t)$. This gives 
 \begin{equation}
 \rho(t) \propto L_1  (1+ \log (\Gamma_1t)).
 \end{equation} 
 At the same time, the center of mass $R(t)$ exhibit a diffusive motion 
 $R(t)=\sqrt{ D_2(t) t}$, with a slightly time dependent diffusion constant 
 $ D_2(t) \approx (U^2/B) (L_1/ \log (\Gamma_1 t))$. This small time 
 dependence of $D_2(t)$ comes from the fact that the frequency of the 
 collisions between the two particles decreases as the pair size grows. 
 This diffusion stops when $R(t) \approx L_2 \approx (U/B)^2 L_1^2$, where 
 TIP localization occurs.

\subsection{Level curvature} \label{b}

  Useful, though indirect information for the TIP propagation at 
 scales $L \leq L_1$ can be found in Ref.~\cite{ap} where the sensitivity 
 of the TIP levels $E_A$ to a change of boundary  conditions is given. 
 Detailed numerical calculations confirming the predictions of Ref.~\cite{ap} 
 are given in the last paper of this series~\cite{ww}. For a ring  
 threaded by an AB-flux $\Phi$, the TIP curvature $C_2(E) \equiv \sum_{A}  
\frac{\partial^2 E_{A}}{\partial\Phi^2} \delta (E-E_A)$ is given by the 
 expression: 

 \begin{equation}
 C_2(E) \approx g_1 \frac{\Delta_1}{\Delta_2} - (g_1-1) g_2 (U) 
 \frac{\Delta_1}{B}.
 \end{equation}
 $g_1$ and $\Delta_1$ are respectively the one particle conductance and 
 mean level spacing. $g_2(L,U)$ can be understood as an interaction-assisted TIP 
 conductance~\cite{imry} of order $\Gamma_{\alpha\beta} (L,U) / \Delta_2(L)$, 
 where $\Delta_2 (L)$ is the spacing of the free boson levels directly 
 coupled by the interaction. The above expression implies that the effect of 
 $U$ strongly depends on $L$. When $L \ll L_1$, $g_1-1 
 \approx g_1 $, and $C_2$ is mainly given by a kinetic one particle 
 term $g_1 \Delta_1 /\Delta_2$ {\it reduced} by a small correction 
 proportional to $g_2 (L,U)$. This means that for $L \ll L_1$ 
 (ballistic one particle regime) the easy propagation due to kinetic terms 
 is not yet strongly affected by the one particle quantum interference. In 
 this case, the presence of the second particle defavors the propagation of 
 the first and the interaction slightly reduces $C_2$, as confirmed in 
 Ref.~\cite{ww}. When $L\geq L_1$, the one particle transport is suppressed by 
 the quantum interferences (Anderson localization) and the term proportional 
 to  $U$ in $C_2$ {\it changes its sign}, since $g_1-1 \approx -1$ up to 
 exponentially small corrections. Thus, TIP transport is {\it favored} 
 by the interaction, the presence of the second particle leading to a 
 decoherence of the localizing quantum interferences of the first. $C_2$ is 
 of order $g_2$ for the few TIP states re-organized by the interaction when 
 $L \gg L_1$. From the behavior of the TIP curvature, one should expect a 
 change of the role of $U$ for the  TIP quantum motion as a function of 
 time: starting from a localized wave packet, the exploration on a scale 
 $L\leq L_1$ for $t<t_1$ should require a {\it longer} time with interaction 
 than without. But without interaction, the exploration 
 of scales $L \gg L_1$ is forbidden by Anderson localization, while it becomes 
 possible in the presence of interaction.  An important issue is then to 
 know what is the time scale $t_2 - t_1$ required for this exploration, 
 before the pair itself gets localized when $t\approx t_2$. According to 
 Ref.~\cite{fmgp},  $ t_2 - t_1 = \sqrt{L_2/D_2(t_2-t_1)}$. However, one should 
 have in mind some further results obtained for strictly on site interaction 
 and strictly one dimension. 

 \subsection{Multifractality} \label{c}
  
 The interaction matrix elements 
 $$ 
 \langle\alpha\beta|H_{\rm int}|\gamma\delta\rangle  
 = 2U \sum_{n=1}^L \Psi_{\alpha}^* (n) \Psi_{\beta}^* (n) 
 \Psi_{\gamma} (n) \Psi_{\delta} (n),$$ 
 (with $\Psi_{\alpha}(n)=\langle n|\alpha\rangle$ ) 
 defines a measure of the free boson states  $|\gamma\delta\rangle$ 
 coupled to a given $|\alpha\beta\rangle$. It was shown in Ref.~\cite{wp} 
 that this measure is multifractal. In contrast to earlier assumptions, the 
 density $\nu_{\rm eff}(L)$ of free boson states $|\gamma\delta\rangle$ 
 effectively coupled to $|\alpha\beta\rangle$ is much weaker than the 
 total density $\nu_2 (L)$ of free boson states. This is not surprizing 
 when there is no disorder: only free boson states of same total momentum 
 are directly coupled, and $\nu_{\rm eff} = \nu_1 \propto L$ and not $\nu_2 
 \propto L^2$. When disorder is switched on, the momentum is no longer a good 
 quantum number, and one gets $\nu_1 < \nu_{\rm eff} < \nu_2$. More precisely, 
 if one wants to estimate $\Gamma_{\alpha\beta}$ using the Fermi golden rule, 
 one needs the {\it effective} density of states $|\gamma\delta\rangle$ 
 coupled by the second moment ($q=2$) of the interaction matrix elements. 
 The measure in the TIP Hilbert space of the support of this 
 set of states is neither $d=1$ (as for the clean case) nor $d=2$ (as for 
 the chaotic $d=0$ case), but a fractal dimension $1 < f(\alpha (q=2)) 
 \approx 1.75 < 2$. This gives a reduced effective  density  $\nu_{\rm eff} 
 \propto L^{f(\alpha(q=2))} $. The measure is multifractal, since this density 
 depends on the considered $q^{th}$ moment of the coupling term. A direct 
 implication of this multifractal character is that the lifetime of the 
 free boson states is longer than $\Gamma_1$ for $L\approx L_1$. One should 
 expect that the diffusion law obtained ignoring multifractality 
 underestimates the time $t_2-t_1$ for the particles to propagate between 
 $L_1$ and $L_2$. The multifractality of the set of directly coupled free 
 boson states should mean a {\it very slow} interaction induced 
 delocalization.    
  
 \subsection{Weak critical chaos}\label{d}

 The TIP spectral fluctuations also lead us to expect very slow dynamics 
 before localization. For a given $L$, the TIP spectrum has Poisson 
 statistics in the limits where either the disorder $W$ or the interaction 
 $U$ are too weak or too strong. In the limit of the clean system $(W=0)$, 
 this is due to the fact that the effective density of coupled free boson 
 states is $\nu_1 \propto L \ll \nu_2 \propto L^2$. For $W \gg 1$, one has 
 $L_1 \ll L$ and the small  part of TIP levels being reorganized 
 by $U$ is totally hidden behind the main part of the non reorganized 
 spectrum, corresponding to free boson states with $|n_{\alpha}-n_{\beta}| 
 \geq L_1$ and which remain eigenstates when $U$ is switched. For $U \ll 1$, 
 the TIP states are basically the free boson states of energy 
 $\epsilon_{\alpha} + \epsilon_{\beta}$, i.e. almost uncorrelated 
 TIP-levels. For $U \gg 1$, the levels have again an energy 
 $\epsilon_{\alpha} + \epsilon_{\beta}$ (duality property explained 
 in the next paragraph). In the plane $(U,W)$, inside those mentioned 
 Poisson limits, in a domain centered around 
 $L\approx L_1$ and $ U \approx 1$, the TIP spectrum becomes more rigid, 
 though less rigid than a Wigner-Dyson spectrum associated to quantum chaos. 
 The spectral rigidity saturates~\cite{wwp} to an intermediate rigidity 
 between Poisson and Wigner-Dyson. This rigidity is however not arbitrary, 
 but has a {\it universal} character shared by many one particle `critical' 
 systems~\cite{bogomolny}, such as the Anderson model at the mobility edge, a 
 mixed system where integrable and chaotic trajectories coexist or a 
 pseudo-integrable billiard where all trajectories belong to a surface of 
 genus larger than one. By the term ``weak critical chaos'', 
 we mean that under certain circumstances ($U\approx 1$ and $L\approx L_1$), 
 the TIP system belongs to the same critical universality class than those 
 one particle systems, at least as far as the spectral statistics are 
 concerned. The interaction can never drive the TIP system towards a 
 stronger chaos, i.e, towards full quantum chaos with Wigner-Dyson 
 statistics.

 \subsection{Duality}\label{e}

 The Hamiltonian without interaction is diagonal in the basis of the 
 free boson states. As pointed out by Ponomarev and 
 Silvestrov~\cite{ponomarev}, there is an eigenbasis appropriate 
 when $U \rightarrow \infty$, having the same energies $\epsilon_{\alpha} 
 +\epsilon_{\beta}$ than the free boson states. Indeed, when 
 $U\rightarrow \infty$, one has just to solve a non interacting problem 
 with new boundary conditions. Since particles cannot be on the same site, 
 we define $L(L-1)/2$ hard core boson states $|hc\rangle$ of components 
 $ \langle n_1 n_2|hc\rangle $ given by 
  \begin{equation}
 \frac{1}{\sqrt{2}}\left[\Psi_{\alpha}(n_2) 
 \Psi_{\beta}(n_1) - \Psi_{\alpha}(n_1) \Psi_{\beta}(n_2)\right]
 \frac{n_2 - n_1}{|n_2-n_1|}.
 \end{equation}
 A hard core 
 boson state  is just a $2 \times 2$ antisymmetric Slater determinant 
 resymmetrized by the factor $(n_2 - n_1)/|n_2-n_1|$. To complete 
 this basis of the symmetric TIP Hilbert space spanned by $L(L+1)/2$ 
 states, we add $L$ molecular states $|nn\rangle$ of energy 
 $2V_n+U$, $V_n$ being the random potential of the $n^{th}$ site. This 
 set of molecular states forms a small sub-band which goes to very large 
 energies when $U \rightarrow \infty$. For the main sub-band of hard core 
 bosons states, centered around $E=0$ as the free boson states, there is a 
 duality relation~\cite{wwp,ponomarev} $U \leftrightarrow A /U$, 
 mapping the behavior at small $U$ onto the behavior at large $U$. 
 $A=\sqrt{24}$ for $E=0$. One finds that the coupling terms between 
 the free boson states are given by 
 $$
 2U \sum_n \Psi^*_{\alpha}(n) \Psi^*_{\beta}(n)\Psi_{\gamma} (n) 
 \Psi_{\delta} (n),$$ 
 while the coupling terms between the hard core boson states are a 
 sum of terms like 
 $$
  \sum_{n,n',n''} \frac{
 \Psi^*_{\alpha}(n) \Psi^*_{\beta}(n)\Psi_{\gamma} (n') \Psi_{\delta} 
 (n'')}{U+2V_n - E},
 $$ 
 with various combinations of $n',n''=n \pm 1$. The duality is obtained 
 neglecting the difference between $n$ and $n'$ and assuming that 
 $U+2V_n-E \approx U$. This gives a  strong dependence of the TIP 
 dynamics on the initial wave packet. For large $U$, the 
 ($|hc\rangle \cup |nn\rangle$) form an eigenbasis, the motion of 
 a wave packet located at the sites $0$ and $\rho_0$ for $t=0$ is 
 given by the time dependent wave function: 
 $$
 \psi_{n_1,n_2}(t)=\sum_{S=(hc,nn)} C^*_{n_1,n_2,S} C_{0,\rho_0,S} 
 e^{-iE_St},
 $$
 where the summation $S$ goes over the states $|hc\rangle$ and $|nn\rangle$ 
 and $C_{n_1,n_2,S}= \langle n_1,n_2 | S \rangle $.  
 For two particles put on the same site $n$ at the beginning (i.e. with an 
 energy $U+2V_n$), we mainly probe the molecular state $|nn\rangle$ 
 when $U$ is large: the two particles stay on the same site for 
 a very large time. For smaller values of $U$, this initial wave packet  
 starts to probe the  molecular state $|nn\rangle$ before decaying onto 
 the neighboring hard core bosons states of energy of order $U$. 
 To eliminate the time motion associated with the molecular states and 
 to see the time motion associated to the hard core boson states, which 
 only exhibit the duality property, one has to start with 
 two particles put close to each other on two {\it different} sites 
 $0$ and $\rho_0$.

\section{Asymptotic TIP localization and finite size effects \label{L2}}
 
\begin{figure}[tbh]
\epsfxsize=7cm
\epsfysize=8.5cm
\epsffile{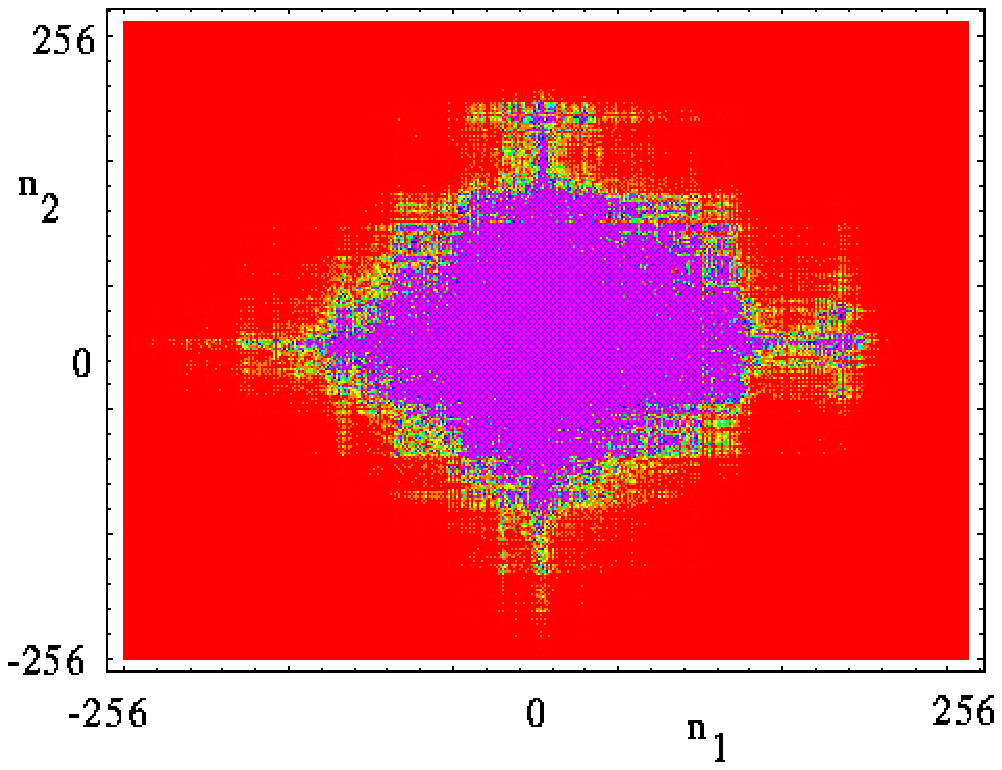}
\epsfxsize=7cm
\epsfysize=8.5cm
\epsffile{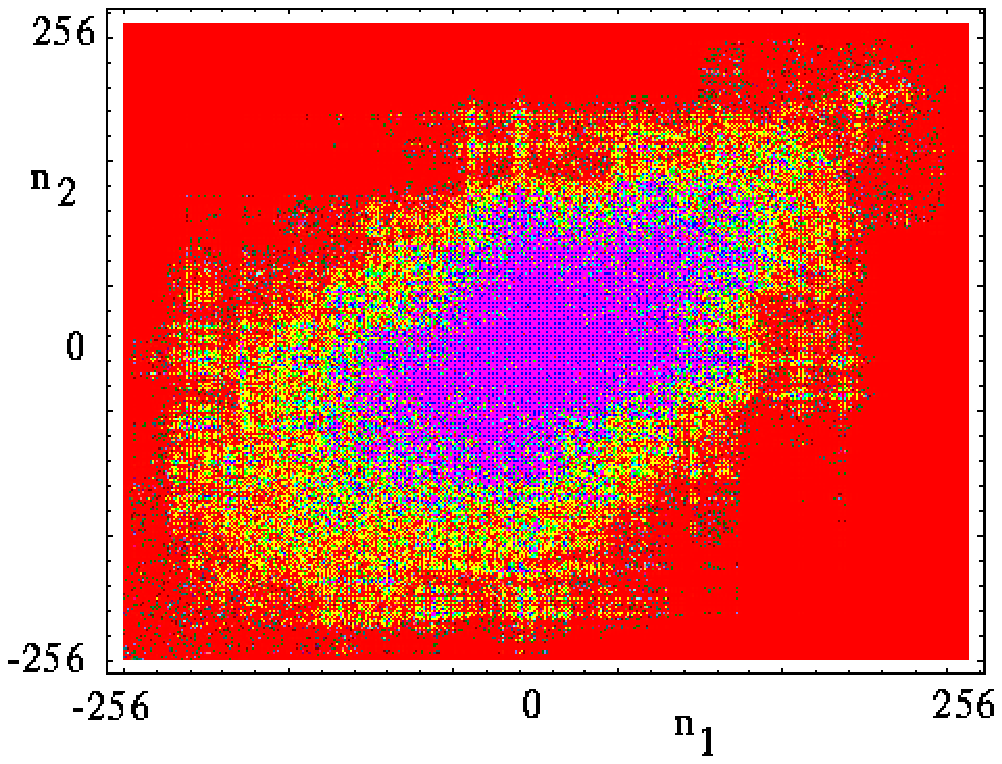}
\vspace{3mm}
\caption[fig2]{\label{tapis1} $|\psi_{n_1,n_2}(t=5.10^4)|^2$  
for two particles put on sites $0$ and $\rho_0=1$ at $t=0$. 
$L=512$ and $L_1=16$. Up: $U=0$.  Down: $U=1$. }
\end{figure}

\begin{figure}[tbh]

\epsfxsize=6.5cm
\epsfysize=6.5cm
\epsffile{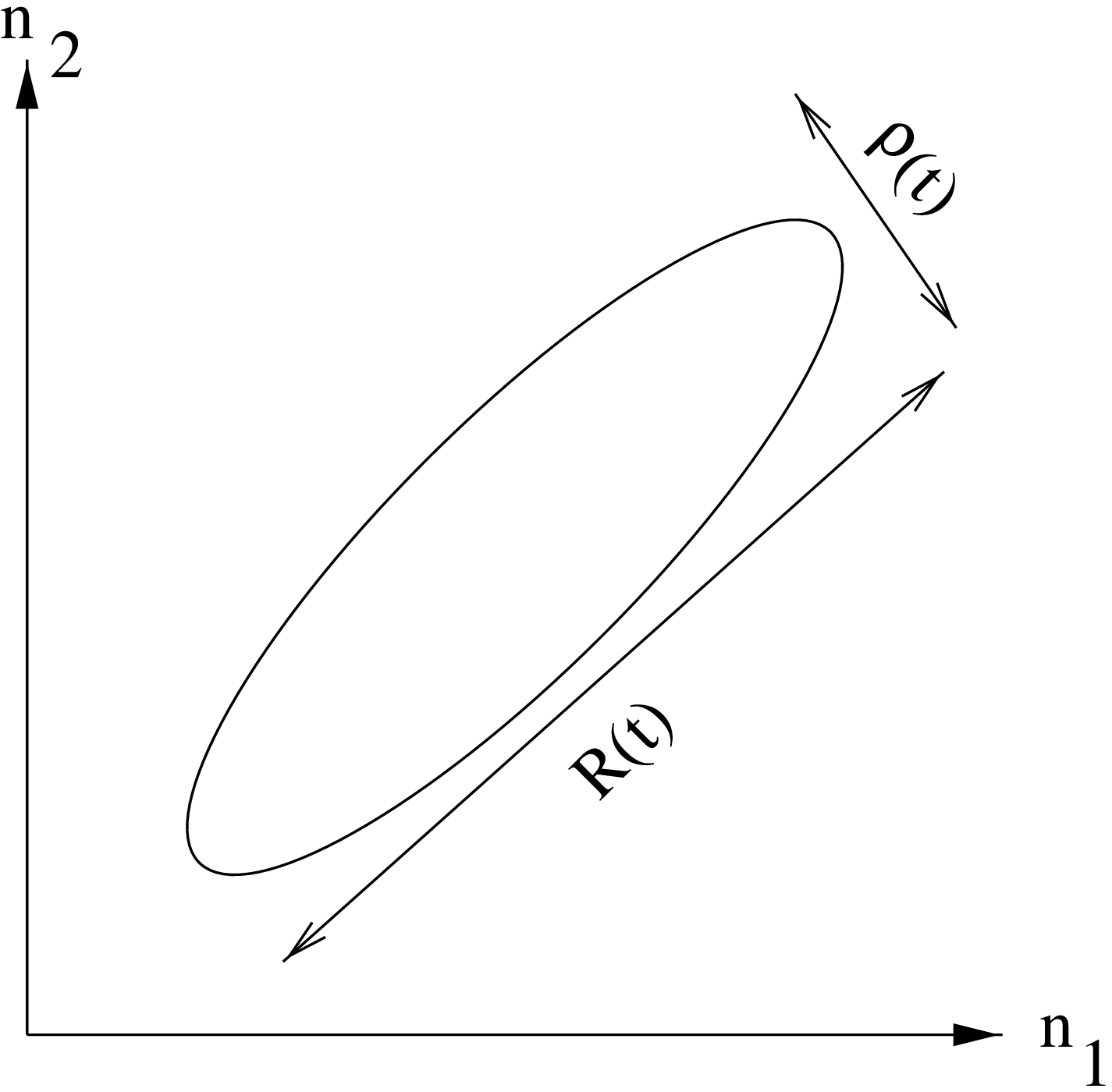}
\epsfxsize=7cm
\epsfysize=8.5cm
\epsffile{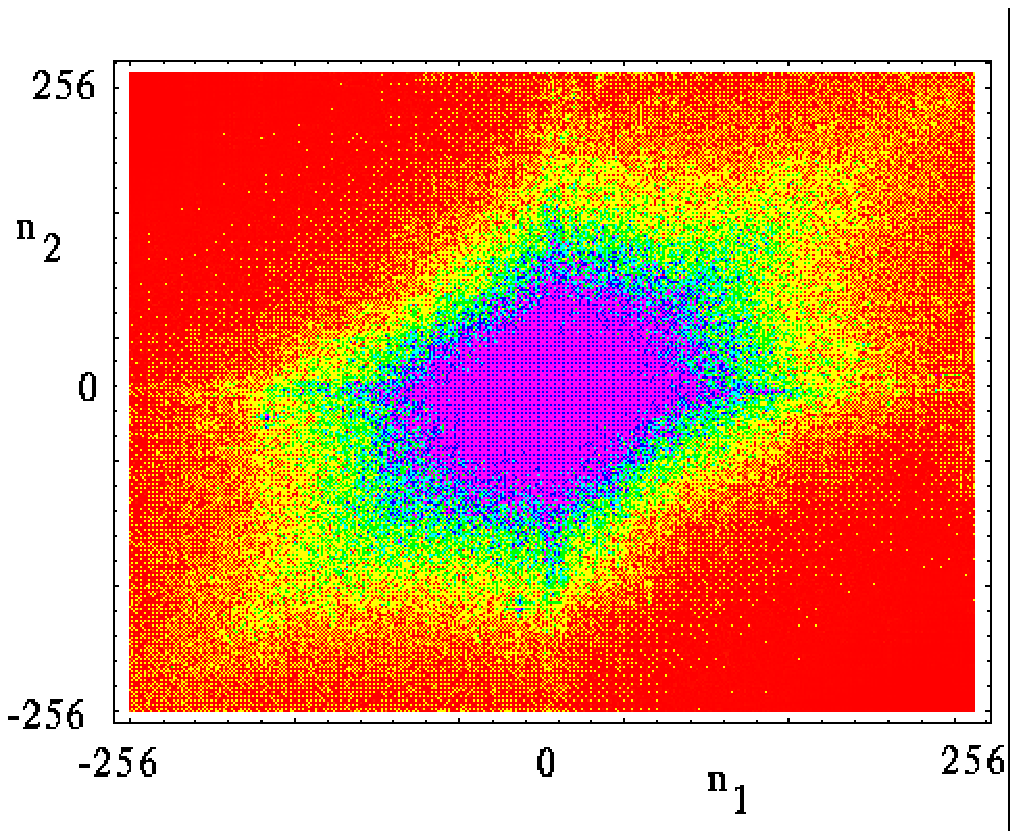}
\vspace{3mm}
\caption[fig2]{\label{tapis2} Up: Scheme of the TIP delocalization effect: 
$|\psi_{n_1,n_2} (\infty)|^2$ is concentrated in an ellipse 
corresponding to a center of mass $R$ delocalized by the interaction 
on a scale larger than the pair size $\rho$. Down: same as in 
Fig.~\ref{tapis1} for $U=1$, $L_1=36$ after having averaged over 20 samples}
\end{figure}

 The role of the repulsive interaction $U$ on the quantum motion is 
shown in Fig.~\ref{tapis1} and Fig.~\ref{tapis2}, illustrating 
the TIP delocalization effect~\cite{s} in a strongly disordered chain 
for $U=1$. This effect is a consequence of the mixing by the 
interaction of free boson states close in energy, delocalizing 
the TIP system in the free boson basis. Since the one body states are 
localized, this delocalization in the free boson basis also means 
delocalization in real space. The initial condition corresponds to 
$\rho_0=1$. We have taken the site potentials $V_0=V_1=0$, in order 
to benefit by the large density of TIP states for $E\approx 0$.

 In Fig.~\ref{tapis1} and Fig.~\ref{tapis2}, the rainbow color code 
indicates on a linear scale the small values of $|\psi_{n_1,n_2} 
(t=5.10^4)|^2$ in red up to the large values in violet. In the upper 
 Fig.~\ref{tapis1}, one can see how the two particles are confined by 
the random potential without interaction, in a localization domain which 
is very quickly reached (typically for $t\approx 200$). In the lower 
Fig.~\ref{tapis1}, $U=1$ and the center of mass becomes delocalized 
as sketched in the upper Fig.~\ref{tapis2}. TIP localization after 
ensemble averaging is shown on the lower Fig.~\ref{tapis2} for 
$U=1$. This TIP ellipsoidal localization domain is reached and stops 
to spread after a considerably larger time (typically for $t\approx 5.10^4$). 
For a given sample, one can see that $|\psi_{n_1,n_2}(t=5.10^4)|^2$ does 
not homogeneously fill the ellipse, and is characterized by large 
fluctuations, mainly near the border of the ellipse. These fluctuations 
are somewhat similar to those characterizing the interaction matrix 
elements coupling the free boson states (see Fig.~1 of Ref.~\cite{wp}). 
As we have checked, $\psi_{n_1,n_2}(t)$ develop an anisotropic 
multifractal behavior when the interaction assisted propagation begins 
to dominate the dynamics.   

 To study the spreading $R(t)$ of the center of mass and the size 
$\rho(t)$ of the pair, we use the following functions:
$$
R(t) = \left(\sum_{n_1,n_2} |\psi_{n_1,n_2}(t)|^2 \frac{(n_1-\bar{n}_1\ 
+ n_2-\bar{n}_2)^2}{2}\right)^{1/2},
$$
$$
\rho(t) = \left(\sum_{n_1,n_2} |\psi_{n_1,n_2}(t)|^2 
\frac{(n_1-\bar{n}_1 - n_2+ \bar{n}_2)^2}{2}\right)^{1/2},
$$
where $\bar{n}_{1,2} =\sum_{n_1,n_2} |\psi_{n_1,n_2}(t)|^2 n_{1,2}$. 
We have checked that the disorder average of $\bar{n}_{1,2}$ does not 
depend on time.

\begin{figure}[tbh]

\epsfxsize=7cm
\epsfysize=7cm
\epsffile{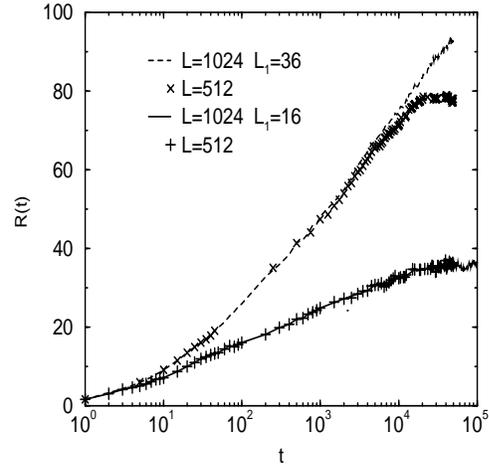}
\vspace{3mm}
\caption[fig3]{\label{bord} Finite size effects for $U=1$. 
$R(t)$ for $L_1=16$ and $36$ for two sizes $L=512$ and $1024$.}

\end{figure}

 Before going further, we first check that the localization effects are 
strong enough so that the TIP motion that we study corresponds to the 
dynamics of the infinite chain, and not of a finite chain with boundary 
effects. Indeed, for a too weak disorder, fast fronts of the wave function 
could propagate~\cite{samjm} up to the boundaries and be reflected. This will 
affect the long time behavior, as the reflected waves enhance the 
localization of the center of mass. The enhancement factor for 
$R(t)$ would be underestimated. Since our numerical 
algorithm allows to study systems of size up to $L=1024$ for $10^6$ 
steps of time, we compare in Fig.~\ref{bord} the motion for 
$L=512$ and $L=1024$ and for $L_1=16$ and $L_1=36$. 

\begin{figure}[tbh]

\epsfxsize=7cm
\epsfysize=7cm
\epsffile{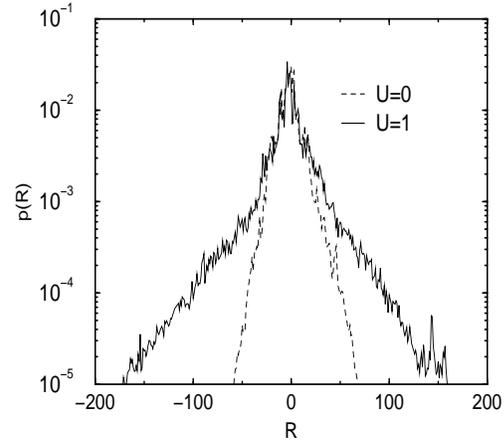}
\caption[fig4]{\label{p(R)} TIP delocalization effect. Probability 
distribution $p(R(t=5.10^4))$ for a single sample with $L_1=16$, 
$L=1024$, $U=0$ (dashed line) and $U=1$ (thick line).}
\end{figure}

For $L_1=16$, $R(t)$ saturates at the same value $L_2\approx 36$  
when the size $L=512$ is doubled, while $R(t)$ has strong finite 
size effects for $L_1=36$ above $t \approx 10^3$ when $L=512$. The 
study of TIP delocalization in the infinite chain can be investigated 
when $L=512$ only for a large disorder, where both $L_1$ and $L_2$ are 
relatively small. Since $L_2$ is reached after a very long time, our 
numerical method is not convenient to study how $L_2$ depends on 
$L_1$ and $U$. We just show on Fig.~\ref{p(R)} the probability density 
$p(R) \equiv \sum_{n_1+n_2=R} |\psi_{n_1 n_2}(t=5.10^4)|^2$ for $U=0$ 
and $U=1$ when $L_1=16$. This proves that the center of mass is 
indeed exponentially localized over a length $L_2 \approx 2 L_1$, without 
finite size effect. For larger values of $L_1$ we have seen larger effects 
($R(U=1)/R(U=0)\approx 3.5$) at long times, but this only gives 
a lower estimate for the enhancement factor $L_2/L_1$, boundary effects 
being non negligible. 

\subsection*{The different regimes of the quantum motion}

\begin{figure}[tbh]

\epsfxsize=7cm
\epsfysize=7cm
\epsffile{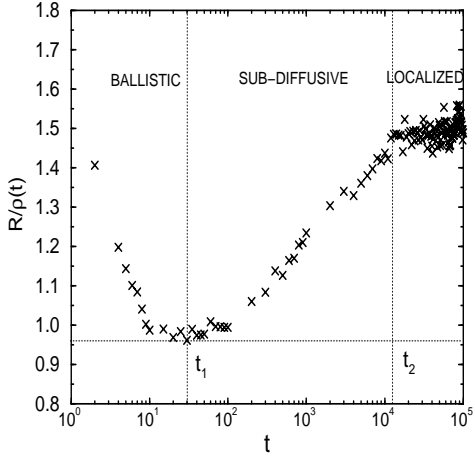}
\epsfxsize=7cm
\epsfysize=7cm
\epsffile{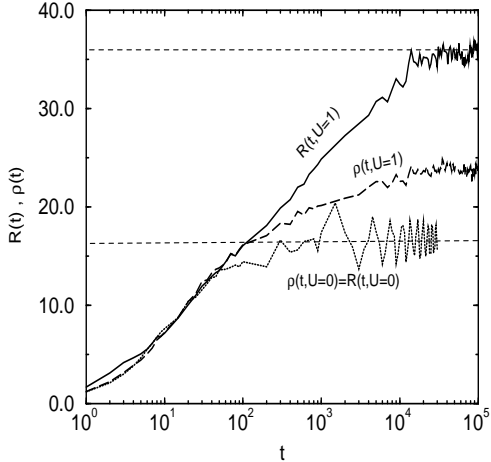}
\vspace{3mm}
\caption[fig5]{\label{R-rho} Single sample with $L_1=16, L=1024$ and 
$U=1$. Up: $R(t)/\rho(t)$. Down: $R(t)$, $\rho(t)$ and $\rho(t)$ for 
$U=0$ and $U=1$.}

\end{figure}

 We now study the intermediary time scales during which the center 
of mass $R(t)$  spreads, before the time $t_2$ where it 
saturates and TIP localization occurs. 
For $U=0$, the aspect ratio of $|\psi_{n_1,n_2}(t)|^2$, defined by  
$R (t)/\rho(t)$, remains equal to one at all times, but for $U\neq 0$, 
the time evolution of this ratio exhibits three regimes (Fig.~\ref{R-rho}), 
delimited by two characteristic time scales $t_1$ and $t_2$. For 
$t \leq t_1$, the repulsive interaction favors $\rho(t)$ and 
defavors $R(t)$. The ratio $R(t)/\rho(t)$ decreases. This is the 
ballistic regime characterizing the length scales smaller than 
$L_1$. The situation is opposite for $t_1<t<t_2$ where 
$L_1$ has been reached and the interaction assisted propagation of the 
center of mass begins, on scales larger than $L_1$. The increase of 
$R(t)$ is now much faster than the increase of $\rho(t)$, and the 
ratio $R(t)/\rho(t)$ increases. $L_2$ is reached at $t=t_2$ where TIP 
localization occurs.

\section{Ballistic propagation and duality}\label{ballistic}

\begin{figure}[tbh]
\epsfxsize=7cm
\epsfysize=7cm
\epsffile{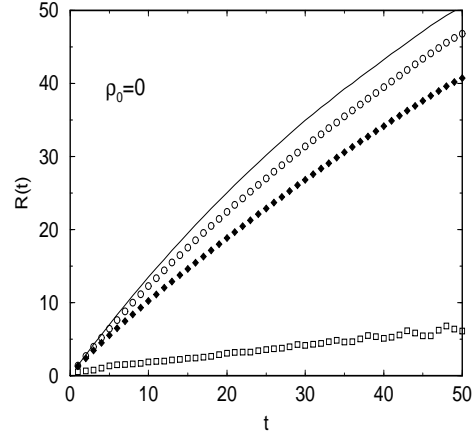}
\epsfxsize=7cm
\epsfysize=7cm
\epsffile{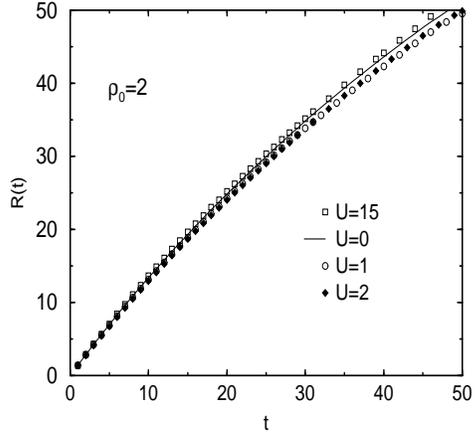}
\vspace{3mm}
\caption[fig6]{\label{t-court} $R(t)$ at small times for different 
values of the interaction. $L_1=L=200$, average over 20 samples. Up:
$\rho_0=0$. Down: $\rho_0=2$.}
\end{figure}

 For $t \leq t_1$ we find that the spreading of the center of mass is 
almost ballistic:
$$ 
R(t) \sim v(U) t^{\mu(U)} {\rm  \ \ with \ \ } \mu(U) \approx 1, 
$$
and that the interaction reduces the increase of $R(t)$. The time evolution 
strongly depends on the initial condition. When the two particles are 
injected on the same site at $t=0$, with an energy of order $U$, the spreading 
of the center of mass is almost suppressed by a too large interaction. 
This is the dynamics associated to the molecular states 
$|nn\rangle$, which do not decay when $U$ becomes very large 
(Fig.~\ref{t-court}).

On the contrary, injecting the two particles at two 
neighboring sites ($0$ and $\rho_0=2$), one can see 
the dynamics associated to the hard core boson states and the consequence 
of the duality relation $U \leftrightarrow 1/U$ between the free bosons 
and the hard core bosons. Since the density of those states exhibits a 
Van Hove singularity at $E=0$ for $U,W=0$, we have also taken 
$V_{0}=V_{\rho_0}=0$. This optimizes the coupling between $\psi(t=0)$ 
and the hard core spectrum. The duality shows up in the 
quantum motion. On Fig.~\ref{t-court}, one can see (i) that when 
$\rho(t=0)=2$, $R(t)$ increases almost linearly as a function of $t$ 
(i.e. that the motion is almost ballistic, (ii) that $U<2$ decreases $R(t)$, 
(iii) that $R(t)$ is very similar when $U=0$ and $U=15$ (duality). 

\begin{figure}[h]
\epsfxsize=7cm
\epsfysize=7cm
\epsffile{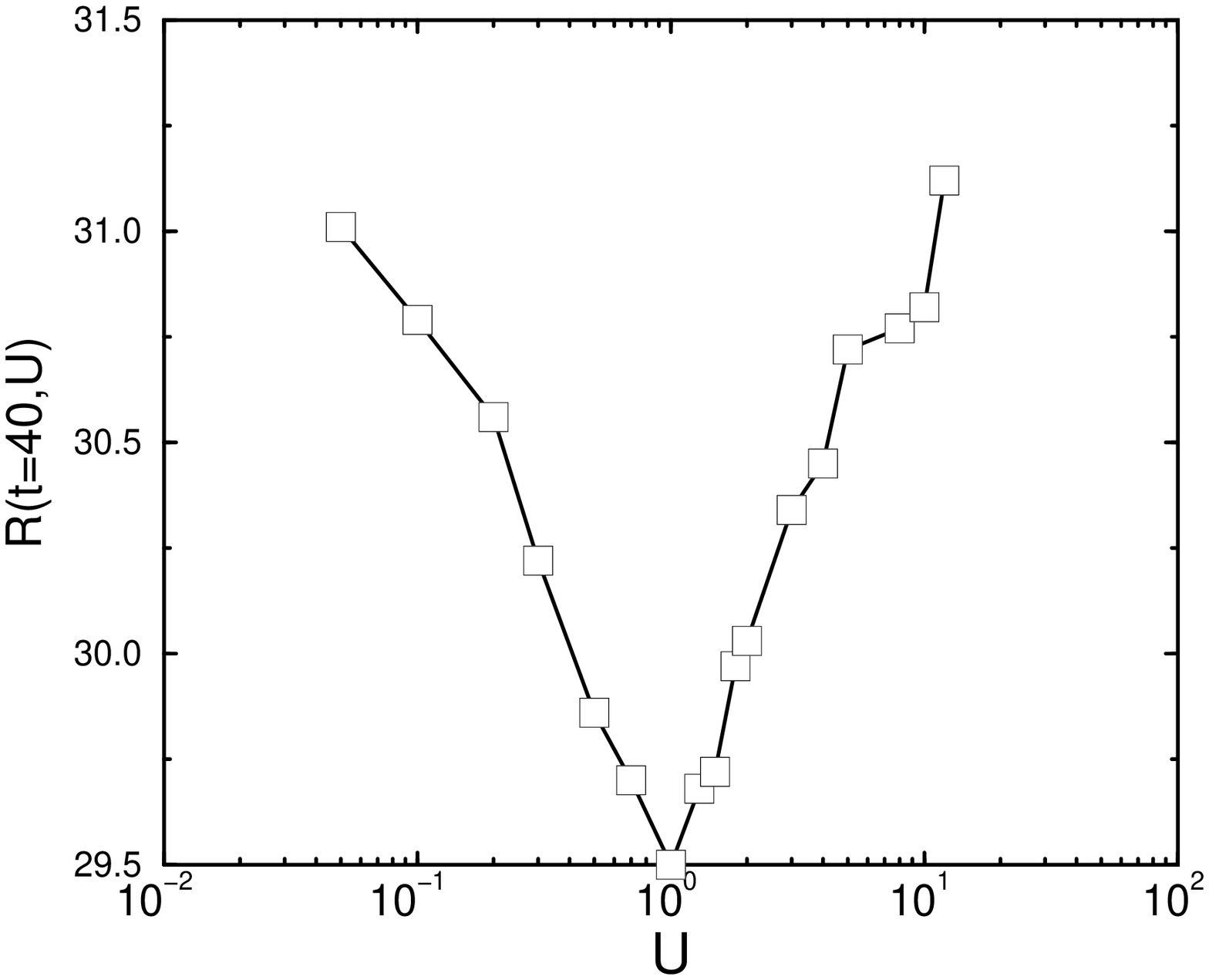}
\epsfxsize=7cm
\epsfysize=7cm
\epsffile{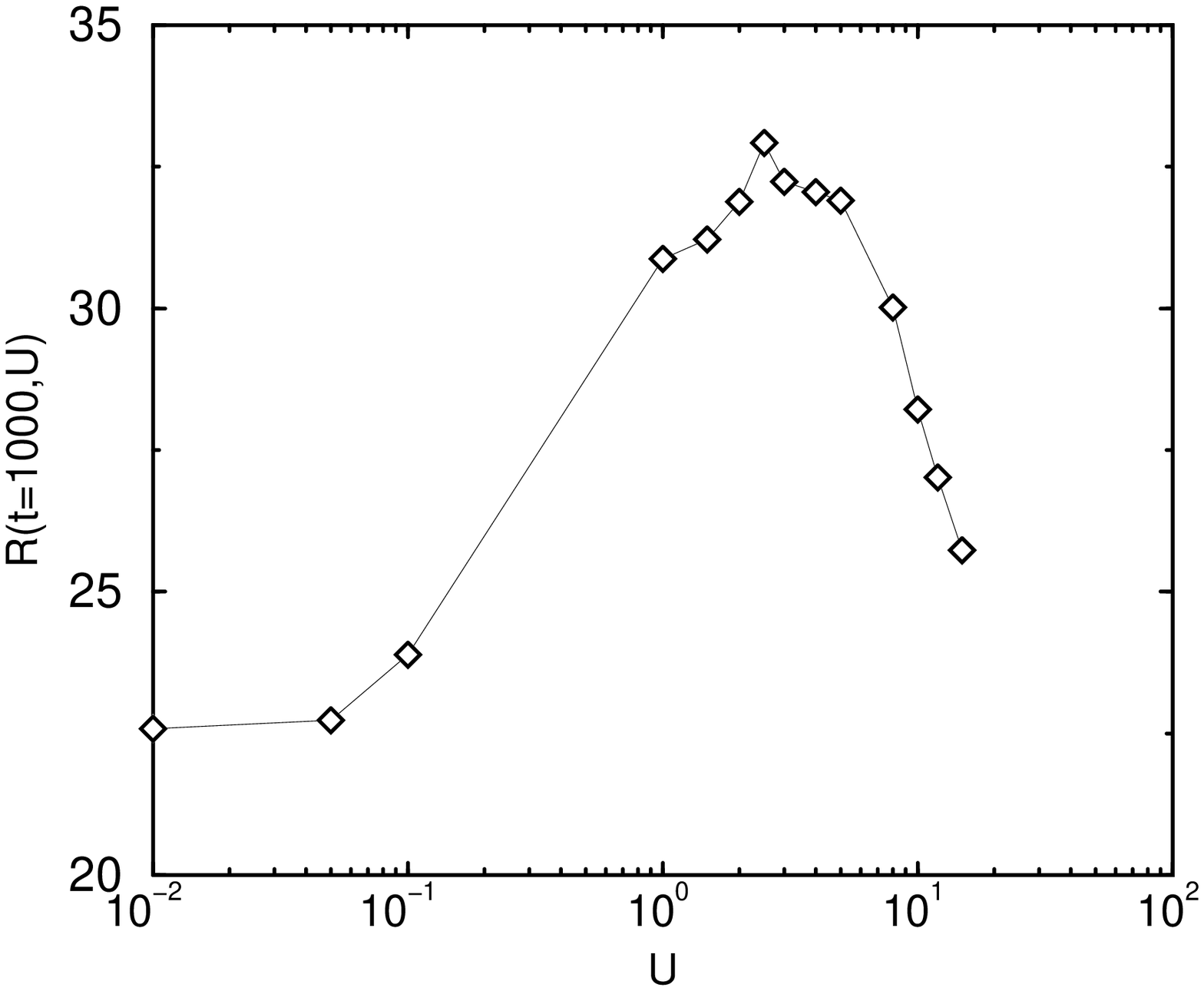}
\vspace{3mm}
\caption[fig7]{\label{dual} Duality and inversion of the role of $U$: 
$R(t)$ averaged over 100 samples. Up: ballistic regime $t=40<t_1$, 
$L=L_1=200$. Down: sub-diffusive regime, $t=1000$, $L=512$, $L_1=36$.}
\end{figure}

 On Fig.~\ref{dual}, the averages over the random potential of $R(t=40)$ 
and $R(t=1000)$ are shown as a function of the strength $U$ of the 
interaction. The upper figure corresponds to the ballistic regime where 
$U$ defavors transport, while the lower figure corresponds to the case 
where $U$ has started to favor transport. The two curves illustrate the 
consequences of the duality between the free bosons and the hard core 
bosons and the inversion of the role of $U$ around $t_1$.

\section{Scaling in the ballistic regime} \label{t_1}
\begin{figure}[tbh]
\epsfxsize=6cm
\epsfysize=5.7cm
\epsffile{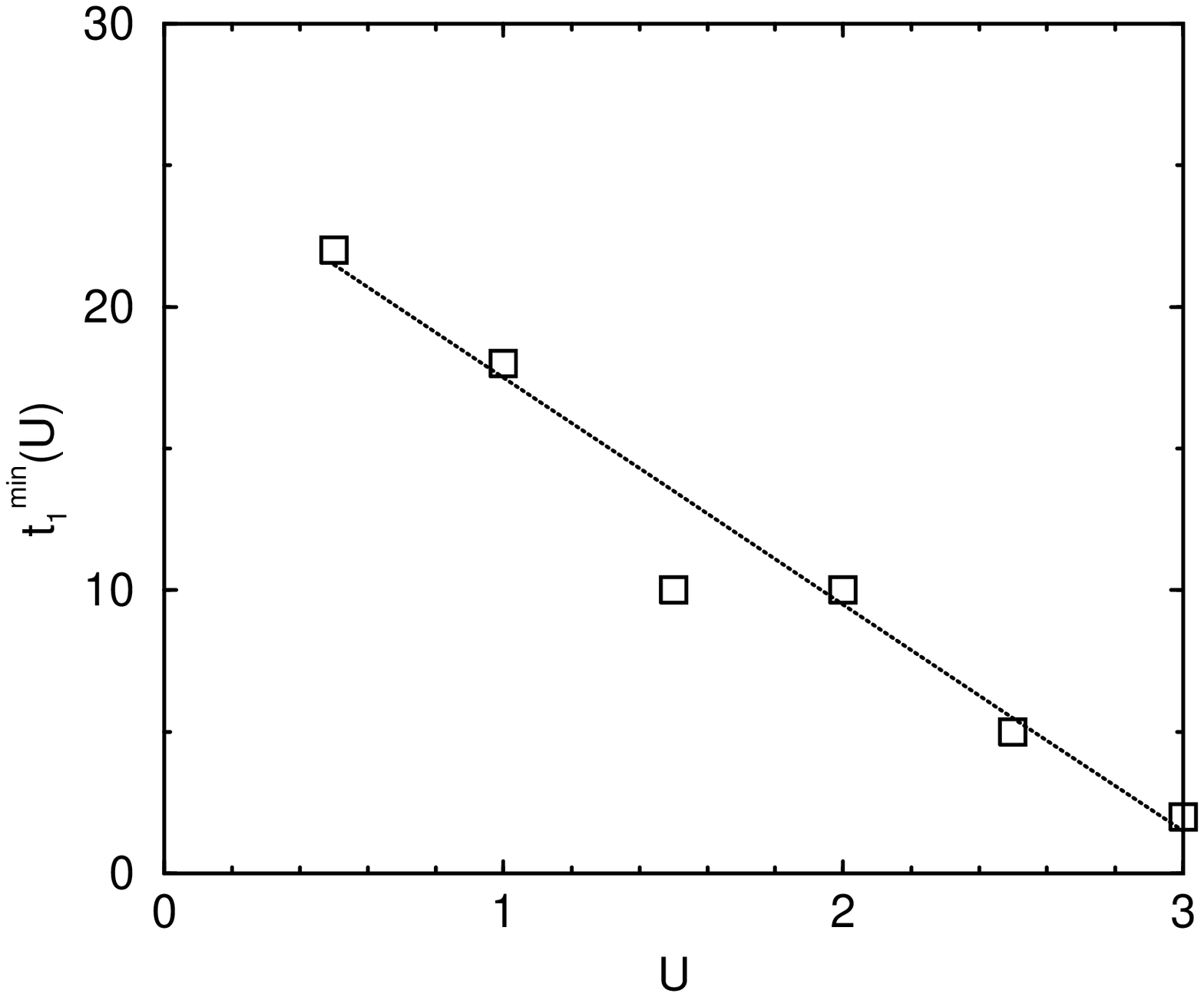}
\epsfxsize=6cm
\epsfysize=5.7cm
\epsffile{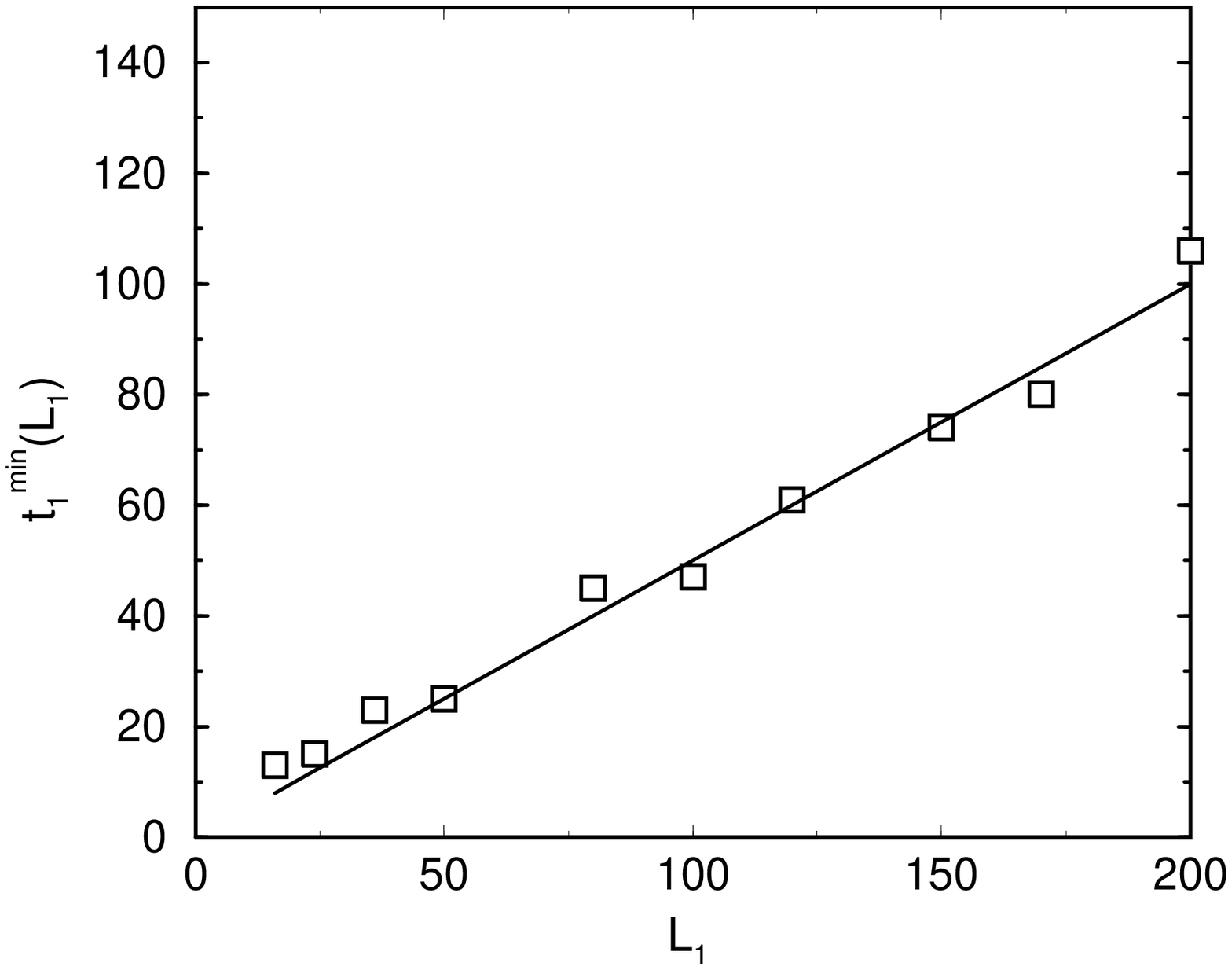}
\epsfxsize=6cm
\epsfysize=5.7cm
\epsffile{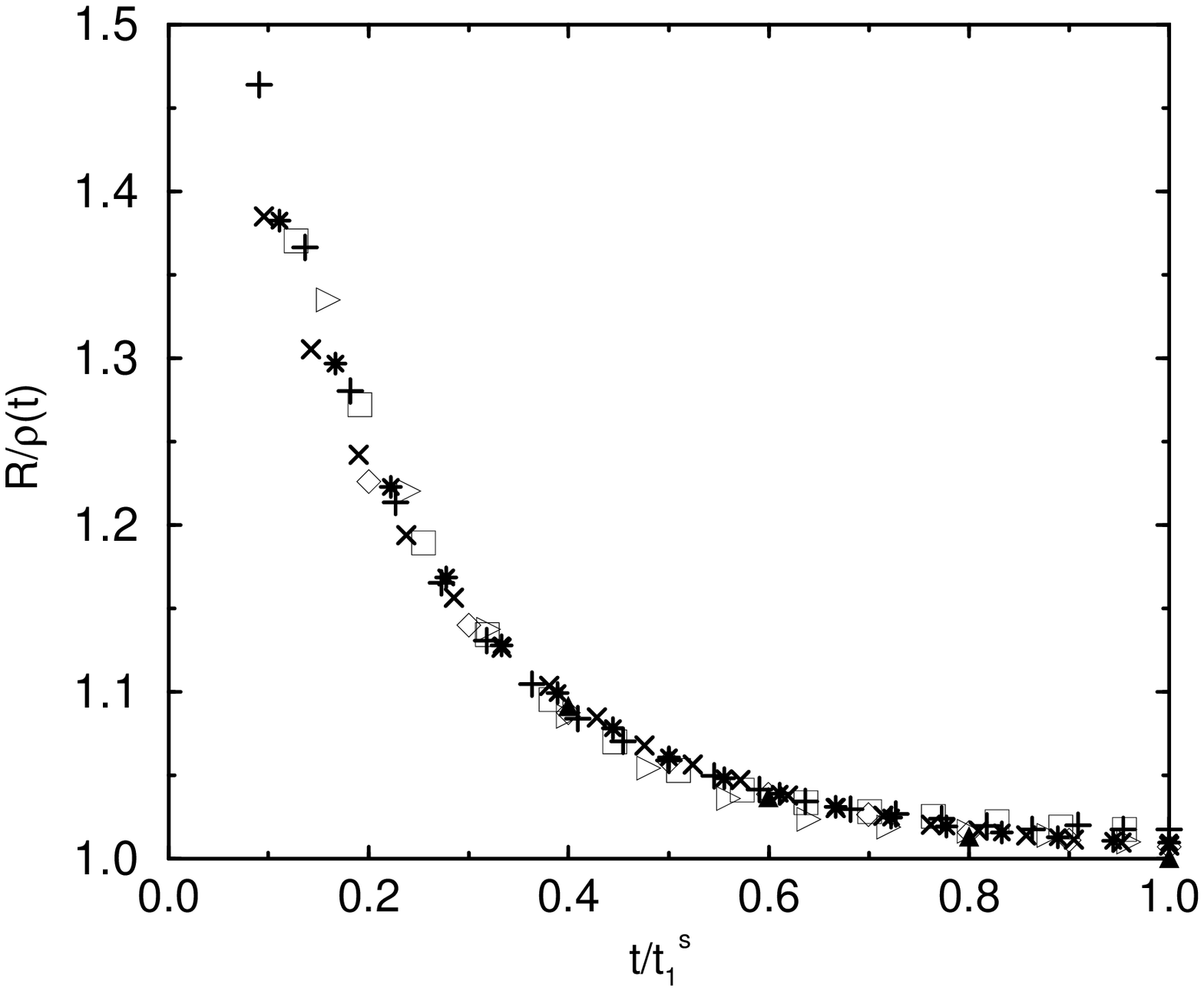}
\vspace{3mm}
\caption[]{\label{t1} Up: $t_1^{\rm min}(U)$ with $L_1=24$, $L=256$ and 200 
samples. The line is a linear fit:
 $t_1 = L_1 - 8 U$. Middle: $t_1^{\rm min}(L_1)$ with $L\approx 3 \times L_1$, 
$U=1$ and 50 samples. The line
is a linear fit: $t_1^{\rm min} = 0.5 L_1$. Down: 
rescaling $R(t/t_1^s)$ with $L_1=24$, $U=0.5$ (pluses), $U=1$ (squares), $U=1.5$
 (diamonds), $U=2$ (full triangles) and $U=1$ $L_1=16$ (triangles), $L_1=36$ (stars), $L_1=50$ (crosses). }
\end{figure}
 To define the characteristic time $t_1$ separating the ballistic and 
sub-diffusive regimes, one can use many criteria. 
For instance, $t_1$ can be defined (i) as the lifetime $\hbar / \Gamma_1$ 
of the free (hard core) boson states, when $U<U_c\approx 2$ ($U>U_c$); 
(ii) as the time $t_1^R$ where the particles reach the one particle 
localisation length $L_1$ when $R (U,t_1^R)=R (U=0, t=\infty)$; (iii) 
as the time $t_1^{\rm min}$ where the minimum of $R (t)/\rho(t)$ is reached 
(see Fig.~\ref{R-rho}); (iv) as the time scale $t_1^{s}$ allowing to map 
the curves $R(t)/\rho(t)$ onto a single scaling curve $R(t)/\rho(t)=
f_s(t/t_1^s)$. The existence of such a scaling and the $L_1$ and $U$ 
dependence of $t_1^s$ are shown on Fig.~\ref{t1}.

 We have checked that the definitions (iii) and (iv) are compatible. 
For $t<t_1$ the motion is essentially ballistic, and we expect that 
$t_1(L_1) \propto L_1$, as for a {\it clean} system of size $L=L_1$ 
(definition (i)) where a term of order $\pm U/L_1^2$ coupled $\nu_1 
\propto L_1$ free boson states.
$$ 
\frac{t_1}{L_1} = f(U) \approx 1 - 0.3 U.$$
The interesting feature of $t_1^{\rm min}$ is that $R \approx \rho $ at 
this time, as when $U=0$. This time where 
there is an inversion of the effect of the interaction should be 
related to the size $L$ where the TIP level curvature~\cite{ww} 
does not depend on $U$. According to Ref.~\cite{ww}, this size is of 
order (but not exactly) $L_1$. 

\section{Very slow delocalization and weak critical 
chaos} \label{sousdif}

 After the ballistic propagation for $t < t_1$, the spreading of the center 
of mass measured by $R(t)$ saturates without interaction. This is due to 
one particle quantum interferences yielding one particle Anderson 
localization. When $U \neq 0$, this saturation is suppressed, but the 
spreading $R(t)$ has now a so slow increase that a logarithmic scale 
for the time $t$ is appropriate. 

\begin{figure}[tbh]

\epsfxsize=7cm
\epsfysize=7cm
\epsffile{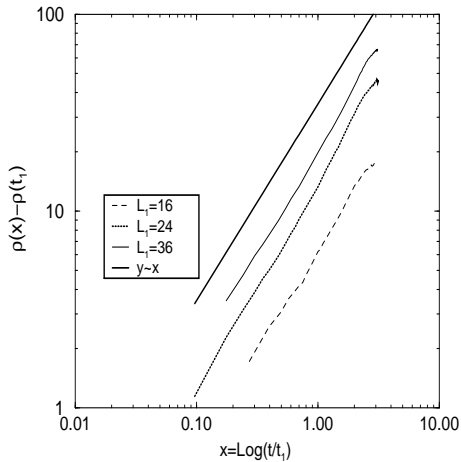}
\caption[fig9]{\label{size}
 Dynamics of the size of the pair: $\rho(x)-\rho(t_1)$ as a function 
of $x=\log(t/t_1)$ in log-log  coordinates for $L_1=16,24,39$, $L=512$ 
and $U=1$.}
\end{figure}

   Let us first consider the increase of the relative separation $\rho(t)$ 
between the two particles when interaction assisted propagation begins. 
As recalled in the subsection~\ref{a}, we expect a behavior given by  
$$ \rho(t) \approx L_1(1+\ln (\Gamma_1 t)),$$
which turns out to be in good agreement with the numerical results. 
Plotting $\rho(x )-\rho(t_1)$ as a function of $x=\log(t/t_1)$, one can 
check on Fig.~\ref{size} the predicted logarithmic behavior.

 The evolution of the center of mass is on the contrary not described by the 
(modified) diffusion law 
$$
R(t) \approx \sqrt{D_2(t) t}, 
$$ 
but has a much slower motion. However, this slow interaction induced 
delocalization is not very surprising since:

(i) The hopping terms induced by the interaction between 
the free boson states (or the hard core boson states) are much 
smaller in one dimension than assumed in the random matrix model 
used to derive the non linear $\sigma$ model of Ref.~\cite{fmgp} 
(subsection~\ref{a}). The interaction matrix is not Gaussian and the 
effective density of coupled free boson 
states is significantly reduced (subsection~\ref{c}). Due to this 
multifractal perturbation, the lifetime of the free boson states 
is much larger than for a more normal (Gaussian) interaction matrix, 
as checked in Ref.~\cite{wp}.

(ii) For $L\approx L_1$, the interaction can only drive weak critical 
 chaos (subsection~\ref{d}). For $L \geq L_1$, 
 the spectrum becomes less rigid, with statistics intermediate between 
 critical statistics and Poisson statistics. Indeed, for $L \geq L_1$, 
 the TIP spectrum becomes a superposition of many states not reorganized 
 by $U$ and  having uncorrelated fluctuations (Poisson) and a small part 
 of states having critical statistics (weak critical chaos). With the 
 chosen initial condition (two particles put close the one to the other at $t=0$) 
 one can argue that we mainly probe the few states with critical 
 spectral statistics. Those critical statistics are associated with slow 
 anomalous diffusion. Let us take one of those billiards with critical 
 statistics: a right triangle~\cite{bogomolny} with smallest angle equal to $\pi/5$ and 
 Dirichlet boundary conditions. For a very long time, the classical 
 trajectories are stable, the system remains on the same KAM torus in 
 phase space, until the corner with angle $4\pi/5$ is reached. At this 
 moment only, the trajectories can escape from the original KAM torus 
 and start to explore other parts of the phase space.  This suggests us 
 a possible analogy between a single particle in the triangle billiard 
 and the two particles in a disordered chain of size $\approx L_1$ with 
 {\it on site interaction}: each particle is trapped in the one particle 
 phase space between the collisions. Only after a collision, the frequency 
 of those collisions being very low and depending on the strength of the 
 random potential, the two particle phase space starts to be explored. This 
 may explain the similarity with a single particle in the triangle. 
 Critical statistics are associated with very slow explorations 
 of the phase space,  and hence of the real space. The fact that the interaction
 can never drive full quantum chaos, but only weak critical chaos, makes 
 likely a very slow interaction assisted propagation. If the motion was 
 isotropic in the plane ($n_1,n_2)$, we should expect anomalous diffusion 
 ($R^{2}(t) \approx t^{\alpha}$) with $\alpha < 1$, as observed for a single 
 particle at the mobility edge in three dimensions, where the spectral 
 statistics are critical.

\begin{figure}[tbh]
\epsfxsize=7cm
\epsfysize=7cm
\epsffile{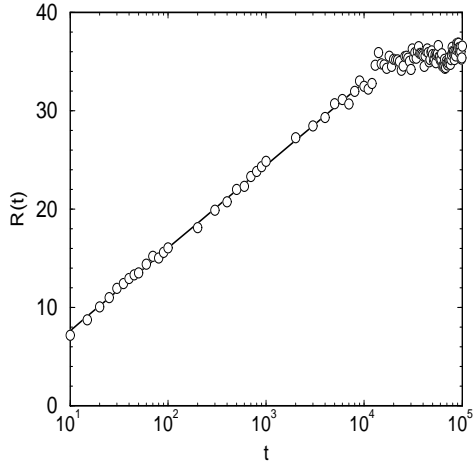}
\vspace{3mm}
\caption[fig9]{\label{slow}
 Slow delocalization of $R(t)$: $L_1=16$, $L=1024$ and $U=1$ }
\end{figure}

 Since the motion is anisotropic in the plane ($n_1,n_2$), we do not find 
 simple anomalous diffusion, but a simple $\log(t)$ behavior: 
\begin{equation}
R(t) \propto \log(t),
\end{equation}
as shown on Fig.~\ref{slow} for $L_1=16$ (see also Fig.~\ref{bord}  
for $L_1=36$). 

 This logarithmic delocalization is not easy to explain, but reminds 
us the classical problem of a random walk in a percolating network. 
Without anisotropy, one gets anomalous diffusion. When one 
introduces anisotropy in the walk, it has been observed~\cite{hambourg} 
that the dynamics is no longer described by a power law 
$R(t) \approx t^{\alpha}$,  but by a $\log(t)$ law. If we order the 
localized one particle states $|\alpha\rangle$ by the location 
$n_{\alpha}$ of the center of their localization domain, from one side 
of the chain to the other, the TIP model in the free boson basis 
$|\alpha \beta \rangle$ is both anisotropic (larger hopping along the 
center of mass direction than along the other direction) and the hopping 
terms $\langle \alpha\beta | H_{int} |\gamma \delta\rangle$ have a 
multifractal measure. One can then argue that the two problems might be 
related, sharing the same $\log(t)$ spreading of $R(t)$.

\section{Conclusion}

In summary, this study of the TIP quantum motion has given some new 
insights on several aspects of the model. There are two length 
scales $L_1$ and $L_2$, and two corresponding time scales $t_1$ and $t_2$. 
For $L<L_1$ ($t<t_1$) the interaction defavors the pair propagation and reduces the 
level curvature. On the other hand, for $L>L_1$ ($t>t_1$), there is a 
very slow interaction assisted pair propagation and the level curvature 
increases as a function of $U$. Our main result is that this delocalization 
is very slow, in qualitative agreement with the concepts of interaction 
induced weak critical chaos and of multifractal hopping terms. Moreover, 
the wavefunction $|\Psi_{n_1,n_2}(t)|$ itself has multifractal features 
visible in Fig.~\ref{tapis1}. With appropriate initial conditions, one 
can observe the symmetry $ U \leftrightarrow A/U $. In 
conclusion, we underline that our results characterize symmetric states 
with purely on site interaction in strictly one dimension. It will be 
interesting to study if they remain valid for longer range interactions, 
or in a quasi-one dimensional limit, or if the behavior predicted from 
the $\sigma$ model approach of Ref.~\cite{fmgp} becomes valid.

\section*{Acknowledgements}
We are indebted to the {\sc Inria Sophia-Antipolis} staff for their 
technical assistance for using their CM-$200$ Connexion Machine. 
We warmly thank Jean-Marc {\sc Luck}, Christian {\sc Vanneste} 
and Dietmar {\sc Weinmann} for 
their constant interest in this work.


\begin{thebibliography}{99}

\bibitem{wp}
X.~Waintal and J.-L.~Pichard, TIP I,  
{\it  Eur. Phys. J.} {\bf B} {\bf 6}, 117 (1998).

\bibitem{wwp}
X.~Waintal, D.~Weinmann and J.-L.~Pichard, TIP II, 
 {\it  Eur. Phys. J.} {\bf B} {\bf 7}, 451 (1999).

\bibitem{ww}
A.~Wobst and D.~Weinmann, TIP IV, to appear in
 {\it  Eur. Phys. J.} {\bf B} (1999).  

\bibitem{bogomolny}
E.~Bogomolny, U.~Gerland and C.~Schmit, to appear in
{\it Phys. Rev.} {\bf B 59} (1999).

\bibitem{shapiro}
B.I.~Shklovskii, B.~Shapiro, B.R.~Sears, P.~Lambrianides and H.B.~Shore, 
{\it Phys. Rev.} {\bf B 47}, 11487 (1993).

\bibitem{detoro}
S.~De~Toro~Arias and C.~Vanneste {\it J. Phys. I France.}{\bf 7},1071 (1997)

\bibitem{s} 
D.L.\ Shepelyansky, {\it Phys.\ Rev.\ Lett.\ } {\bf 73}, 2067 (1994). 

\bibitem{fmgp}
K.~Frahm, A.~M\"uller-Groeling and J-L.~Pichard 
{\it Phys. Rev. Lett.} {\bf 76}, 1509 (1996); {\it Z. Phys.} {\bf B} 102, 
261 (1997).

\bibitem{ap}
E.~Akkermans and J.-L.~Pichard, {\it Eur. Phys. J.} {\bf B 1}, 223 (1998).

\bibitem{imry} 
Y.~Imry, {\it Europhys.~Lett.}~{\bf 30}, 405 (1995).
 
\bibitem{ponomarev}
I.V.~Ponomarev and P.G.~Silvestrov, {\it Phys. Rev.}  {\bf B 56}, 3742 (1997).

\bibitem{samjm}
S.~De Toro Arias and J.M.~Luck, to appear in {\it J. Phys.} 
{\bf A}, cond-mat/9808021. 

\bibitem{hambourg} J.~Dr\"ager, private communication.


\end{thebibliography}
\end{document}